# ContractFuzzer: Fuzzing Smart Contracts for Vulnerability Detection[*][#]

Bo Jiang[†]
School of Computer Science and Engineering
Beihang University
Beijing, China
jiangbo@buaa.edu.cn

Ye Liu
School of Computer Science and Engineering
Beihang University
Beijing, China
franklin@buaa.edu.cn

W.K. Chan
Department of Computer Science
City University of Hong Kong
Hong Kong
wkchan@cityu.edu.hk

## ABSTRACT

Decentralized cryptocurrencies feature the use of blockchain to transfer values among peers on networks without central agency. Smart contracts are programs running on top of the blockchain consensus protocol to enable people make agreements while minimizing trusts. Millions of smart contracts have been deployed in various decentralized applications. The security vulnerabilities within those smart contracts pose significant threats to their applications. Indeed, many critical security vulnerabilities within smart contracts on Ethereum platform have caused huge financial losses to their users. In this work, we present ContractFuzzer, a novel fuzzer to test Ethereum smart contracts for security vulnerabilities. ContractFuzzer generates fuzzing inputs based on the ABI specifications of smart contracts, defines test oracles to detect security vulnerabilities, instruments the EVM to log smart contracts runtime behaviors, and analyzes these logs to report security vulnerabilities. Our fuzzing of 6991 smart contracts has flagged more than 459 vulnerabilities with high precision. In particular, our fuzzing tool successfully detects the vulnerability of the DAO contract that leads to $60 million loss and the vulnerabilities of Parity Wallet that have led to the loss of $30 million and the freezing of $150 million worth of Ether.

[*] This research is supported in part by NSFC (project no. 61772056), the Research Fund of the MIIT of China (project no. MJ-Y-2012-07), the GRF of Research Grants Council (project no. 11214116, 11200015, and 11201114), and the State Key Laboratory of Virtual Reality Technology and Systems.

[#] ContractFuzzer is Open Sourced.
https://github.com/gongbell/ContractFuzzer

[†] All correspondence must be addressed to Bo Jiang. E-Mail: gongbell@gmail.com. Web: http://jiangbo.buaa.edu.cn/

## 1 Introduction

Cryptocurrencies and blockchain technologies have gained huge popularity and attentions in industry and academia in recent years. By the end of 2017, the total cryptocurrency market capital has reached around 600 billion [19]. A cryptocurrency usually adopts a well-designed consensus protocol that is agreed by all participating nodes in its underlying network, and the computing nodes (i.e., the miners) of the network are responsible for recording the state of the network after transactions in a distributed and shared ledger —— a blockchain. The blockchain was originally proposed for value transfer among networked peers without trust [11]. Later, there are many enhanced blockchain platforms supporting smart contracts. One of the most popular ones is Ethereum [22], which enhances the blockchain platform with a Turing-complete programming language, allowing developers to write smart contracts and decentralized applications. The ecosystem of Ethereum is growing rapidly: the number of smart contracts and decentralized applications have increased to more than 2 million on March 2018 [28]. The explosive growth of Ethereum ecosystem shows its potential in incubating killer blockchain applications in the future.

Smart contracts enable building decentralized applications on top of the blockchain consensus protocol so that users can make agreements via blockchain while minimizing trust. They are code running on the blockchain that can define arbitrary rules to control digital assets [22]. Decentralized Applications (DApps) are basically composed of a set of smart contracts as the backend and a set of user interfaces as its frontend. Smart contracts have enabled a wide range of DApps in practice, such as wallets, prediction market, instant messaging, microblogging, crowdfunding, etc. The Ethereum accounts (including both the smart contract accounts and externally own accounts) are now managing 98 million of Ether, which is about 59 million of USD in early 2018 [28].

However, managing so much wealth under smart contracts also makes them attractive targets for attacking by hackers. Indeed, security problem with smart contracts have resulted in serious losses for the blockchain community. The infamous DAO contract bug [17] led to $60 million US loss. The Parity wallet has suffered from two vulnerabilities[32][33]. The first one has resulted in the loss of $60 million, and the second one has frozen more than $150 million in terms of Ether.

There are several reasons that make smart contracts vulnerable to security attacks. First, each execution of a smart contract is dependent on the underlying blockchain platform and the executions of other cooperating smart contracts. Smart contract developers may easily write vulnerable smart contracts if they fail to fully understand the implicit relationships among those smart contracts. Second, the programming languages (e.g., Solidity) and runtime environments are new to many developers, and these tools are still crude. Vulnerabilities may slip into smart contracts when the deficiencies of the tool chain are not well handled by the developers. Last but not the least, the immunity of smart contracts makes updating them after deployment hard to realize. Despite some workaround measures [34], security bug fixes to smart contracts may take a long time to apply, which will expose vulnerable smart contracts to security threats.

In previous work, several smart contract verification tools are proposed to detect security vulnerabilities in smart contracts. However, there are still limitations with them. First, the detection strategy may be imprecise, which can lead to a high rate of false positives, i.e., the detected vulnerabilities are either unlikely to manifest or impossible to exploit. Second, symbolically verifying all possible paths suffer from the path explosion problem, which may also lead to false negatives if only some paths are verified.

In this work, we propose ContractFuzzer, a fuzzing framework to detect security vulnerabilities in smart contracts on Ethereum platform. ContractFuzzer analyzes the ABI interfaces of smart contracts to generate inputs that conform to the invocation grammars of the smart contracts under test. We define new test oracles for different types of vulnerabilities and instrument EVM to monitor smart contract executions for detecting real smart contract vulnerabilities. Then we also showed the practicability and applicability of ContractFuzzer by deploying 6991 real-world smart contracts on our testnet for ContractFuzzer to perform security fuzzing on them. The experimental results showed that ContractFuzzer detected more than 459 vulnerabilities with very high precision, where each of them has been confirmed by our manual analysis.

The main contribution of this work is three-fold. First, to the best of our knowledge, this work proposes the first fuzzing framework for detecting security vulnerabilities of smart contracts on Ethereum platform. Second, it proposes a set of new test oracles that can precisely detect real-world vulnerabilities within smart contracts. Third, we systematically performed fuzzing on 6991 real world smart contracts on Ethereum platform, and ContractFuzzer has identified at least 459 smart contracts vulnerabilities, including the DAO bug and Parity Wallet bug.

The organization of the remaining sections is as follows. In Section 2, we present the preliminaries of smart contract programming and review typical smart contract vulnerabilities. In Section 3, we will define the test oracle to detect vulnerabilities in smart contract. Then, in Section 4, we present the design of our ContractFuzzer framework. After that, we report a comprehensive experimental study to evaluate the effectiveness of ContractFuzzer in Section 5 followed by a discussion on related work in Section 6. Finally, we conclude our work in Section 7.

## 2 A Review of Smart Contracts

In this section, we will briefly review security vulnerabilities of smart contracts studied in this work.

### 2.1 The Basics of Smart Contracts on Ethereum

The state $s$ of a blockchain is a mapping from addresses to accounts. The Ethereum blockchain platform not only supports external accounts (i.e., owned by human) but also smart contract accounts [27], which have balances in terms of Ether and persistent private storage managed by code. Conceptually, Ethereum [22] can be viewed as a transaction-based state machine, where its state is updated on every transaction. Moreover, the validity of the transactions is verified by the consensus protocol of the underlying blockchain platform. A transaction is a message that is sent from one account to another account. It can include binary data (as payload) and Ether. When the target is a smart contract account, its code is executed and the payload is provided as input data.

The executable code of smart contract is bytecode running on the stack-based Ethereum Virtual Machine (EVM). Developers can program smart contracts using Solidity, a high-level programming language [16], which are then compiled into EVM bytecode. Upon creation, each transaction is charged with certain amount of gas to pay for its execution and to avoid malicious code wasting Ethereum resources. When the gas is used up during contract execution, an out-of-gas exception is triggered, which reverts all modifications made to the state of the account in the sense of transaction.

### 2.2 Vulnerabilities of Smart Contracts on Ethereum

The security vulnerabilities of blockchain enabled decentralized applications can happen at the blockchain level, EVM level, and smart contract level. In this work, we are focusing on security vulnerabilities of smart contracts, which we will briefly review in this section. We will follow the vulnerabilities taxonomy of [1] and [9].

**Gasless Send.** The gasless send vulnerability is due to the fact that when using *send* the recipient contract's fallback function will be invoked but with a fixed gas stipend as determined by the EVM. Usually, the gas limit for the fallback function is 2300 when the amount sent is nonzero. As a result, if the recipient contract has an expensive fallback function, the sender of the ether will get an out of gas exception. If such exception is not checked and propagated appropriately, a malicious sender can keep ether wrongfully while seemingly innocent.

**Exception Disorder**. The Exception disorder is due to the fact that Solidity is inconsistent in terms of exception handling, which is dependent on the way contracts call each other. When a contract calls the function of another, it may fail due to different types of exceptions. When such exception occurs, the handling mechanism is determined by how the calls are made. Given a chain of nested calls where each call is a direct call to the function of a contract, when exception occurs, all the transactions will be reverted (including ether transfer). However, for a chain of nested calls where at least one call is made through low-level call methods on address (*address.call()*, *address.delegatecall()*, or *address.send()*), the rollback of the transaction will only stop at the calling function and return false. From that point, no other side effect can be reverted and no throw will be propagated. Such inconsistencies in terms of exception handling will make the calling contracts unaware of the errors happened during execution.

**Reentrancy**. The reentrancy bug is due to the fact that some of the functions are not designed to be reentrant by the developers. However, a malicious contract deliberately invokes such functions in a reentrant manner (e.g., through fallback functions), it may lose ether. The famous "The DAO" attack just made use of this vulnerability to cause $60 million US loss in terms of ether.

**Timestamp Dependency.** The timestamp dependency vulnerability exists when a smart contract uses the block timestamp as part of the conditions to perform a critical operation (e.g., sending ether) or as the source of entropy to generate random numbers. In a distributed system like blockchain, the miner has the freedom to set the timestamp of a block within a short time interval less than 900 seconds [24]. However, if a smart contract transfer ether based on timestamp, an attacker can manipulate block timestamps to exploit the vulnerability.

**Block Number Dependency**. *The block number dependency vulnerability* is like *Timestamp dependency*. It happens when a smart contracts uses the *block.number* as part of the conditions to perform a critical operation (e.g., sending ether) or as the source of generating random numbers. Indeed, both *block.timestamp* and *block.number* are variable that can be manipulated by miners, so they cannot be used as a source of entropy because of the miners' incentive [30]. Moreover, even using the *block.blockhash()* function with *block.number* as parameters for random number generation is still vulnerable either due the execution mechanism of EVM or due to the transparency of the blockchain.

**Dangerous DelegateCall**. The **delegatecall** is identical to a message **call** except that the code at the target address is executed in the context of the calling contract[27]. This means that a contract can dynamically load code from a different address at runtime while the storage still refers to the calling contract. This is the way to implement the "library" feature in Solidity for reusing code. However, when the argument of the delegatecall is set as msg.data, an attacker can craft the msg.data with the signature of a function so that the attacker can make the victim contract to call whatever function it provides. This is exemplified by the outbreaks of the first round of parity wallet vulnerability [32]. As shown in Table 1, at line 6 the Wallet contract contains a delegatecall with msg.data as its parameter. This makes an attacker can call any public function of _walletLibrary with the data of Wallet. So, the attacker calls the initWallet function (defined at line 10) of the _walletLibrary smart contract and become the owner the wallet contract. Finally, he can send the ether of the wallet to his own address to finish the attack. This attack has led to $30 million loss to the parity wallet users.

**Table 1. Dangerous Delegate Call in Parity Wallet Contract**

| 1 | **contract** Wallet{ |
|---|---|
| 2 |    **function() payable** { //fallback function |
| 3 |      **if** (msg.value > 0) |
| 4 |         Deposit(msg.sender, msg.value); |
| 5 |      **else if** (msg.data.length > 0) |
| 6 |         _walletLibrary.**delegatecall**(msg.data); |
| 7 |    } |
| 8 | } |
| 9 | contract WalletLibrary { |
| 10 |    function initWallet(address[] _owners, uint _required, uint _daylimit) { |
| 11 |      initDaylimit(_daylimit); |
| 12 |      initMultiowned(_owners, _required); |
| 13 |    } |
| 14 | } |

**Freezing Ether**. Another type of vulnerable contract is the freezing ether contract. These contracts can receive ether and can send ether to other addresses via *delegatecall*. However, they themselves contain no functions to send ether to other address. In another word, they purely rely on the code of other contracts (via *delegatecall*) to send ether. When the contracts providing the ether manipulation code performs suicide or self-destruct operation, the calling contract has no way to send out ether and all its ether is frozen. The second round of attack on Parity wallet vulnerability is just because many wallet contracts can only rely on the parity library to manipulate their ether. When the parity library was changed to a contract through initialization and then killed by the hacker. All the ether within the wallets contracts relying on the parity library is frozen.

# 3 Defining Testing Oracles for Vulnerabilities of Smart Contracts

In this section, we will define test oracles for detecting each type of vulnerabilities in smart contracts.

## 3.1 Test Oracle for Gasless Send

Within EVM, the *send()* is implemented as a special type of *call()*. So the oracle **GaslessSend** ensures the call within EVM is indeed a *send()* call and that the *send()* call returns an error code of **ErrOutOfGas** during execution. To check a call is a *send()*, we verify whether the input of the call is 0 and the gas limit of the call is 2300.

## 3.2 Test Oracle for Exception Disorder

We define the test oracle **ExceptionDisorder** as follows: for a chain of nested calls (or *delegatecalls*) originated from a root call (or *delegatecall*), if the root call doesn't throw exception while at least one of its nested calls throws exception, we consider the call chain contains exception disorder. In another word, the exception is not properly propagated back to the root call.

### 3.3 Test Oracle for Reentrancy

The test oracle ***Reentrancy*** is defined based on the two sub-oracles. The first sub-oracle is ***ReentrancyCall*** which checks whether the function call A appears more than once within the call chain originated from call A. The second sub-oracle is ***CallAgentWithValue*** that checks three conditions: (1) there is a ***call()*** invocation with value greater than 0 (the amount of Ether to transfer); (2) there are enough gas stipend for the callee to perform complex code execution (i.e., not ***send*** or ***transer***); (3) the callee of the ***call()*** is the agent contract provided by our tool rather than accounts specified by the contract under test. The ***Reentrancy*** test oracle is defined as:

*ReentrancyCall* ∧ *CallAgentWithValue*

We consider the ***Reentrancy*** vulnerability happens when there is a call that invokes back to itself through a chain of calls and the call has sent ether to our agent contract through ***Call()*** with enough gas stipend such that our agent contract can perform the reentrant call again within its fallback functions. In another word, our *ContractFuzzer* flags reentrancy vulnerability only when it can successfully mount a reentrancy attack on the target contract. We will present the design of our agent contract and its attack scenarios in Section 4.3.

### 3.4 Test Oracle for Timestamp Dependency

The test oracle ***TimestampDependency*** is defined on three sub-oracles. The first sub-oracle is ***TimestampOp*** that checks whether the calls within the current contract have invoked the TIMESTAMP opcode during execution. The second sub-oracle is ***SendCall*** which checks whether the call is a ***send()*** call that sends ether to other contract. The third sub-oracle is ***EtherTransfer*** that checks whether the value of a ***call()*** (the amount of Ether to transfer) is greater than 0. To be specific, the ***TimestampDependency*** is defined as:

*TimestampOp* ∧ *(SendCall* ∨ *EtherTransfer)*

To summarize, we consider the ***TimestampDependency*** happens when the current contract has used block timestamp and the contract either has transferred ether during execution.

### 3.5 Test Oracle for Block Number Dependency

The test oracle ***BlockNumDependency*** is similar to ***TimestampDependency*** except that it checks the use of block number rather than the use of block timestamp. It is also defined based on 3 sub-oracles. The first one is ***BlockNumberOp*** that checks whether the calls within the current contract have invoked the NUMBER opcode during execution. The other sub-oracles are the same as that of the ***TimestampDependency***. And the ***BlockNumDependency*** is defined as:

*BlockNumOp* ∧ *(SendCall* ∨ *EtherTransfer)*

To summarize, we consider the ***BlockNumDependency*** happens when the current contract has used block number and the contract either has transferred ether during execution.

### 3.6 Test Oracle for Dangerous DelegateCall

The test oracle ***DangerDelegateCall*** checks whether there is a ***deletegate*** call invoked during the execution of the current contract and that the function called by the delegate call is obtained from the input (e.g., msg.data) of the initial call to the contract. In another word, the test oracle checks whether the contract under test invokes a delegate call whose target function is provided by a potential attacker.

### 3.7 Test Oracle for The Freezing Ether Contract

The test oracle for ***FreezingEther*** checks whether a contract can receive ether and have used ***delegatecall*** during execution but there is no ***transfer/send/call/suicide*** code within the current contract itself to transfer ether to other address. In another word, the ***FreezingEther*** test oracle labels a contract as vulnerable if its balance is greater then zero during execution, but it has no way to transfer ether with its own code (i.e., using call, transfer, and suicide).

## 4 The Smart Contract Fuzzer

In this section, we first give an overview of our ContractFuzzer tool. Then we proceed to present the design of each core component of the tool in detail.

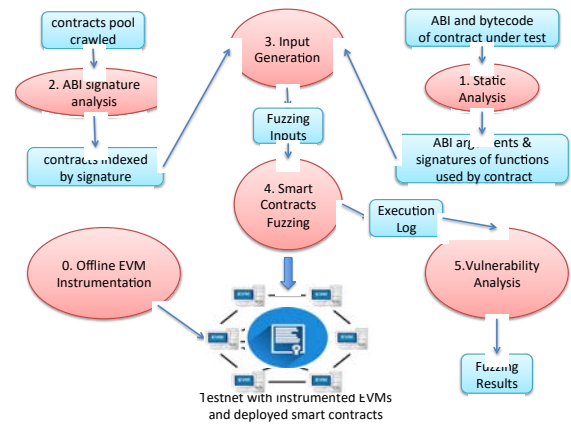

**Fig. 1 Overview of the ContractFuzzer Tool**

### 4.1 An Overview of ContractFuzzer

An overview of ContractFuzzer describing its workflow is shown in Fig. 1 where the major steps are number starting from 0 to 5. The ContractFuzzer tool contains an offline EVM instrumentation tool and an online fuzzing tool. The offline EVM instrumentation process in step 0 is responsible for instrumenting the EVM such that the fuzzing tool can monitor the execution of smart contracts to extract information for vulnerability analysis. We also build a web crawler to extract deployed smart contracts on the Ethereum platform from Etherscan [25] website. Our crawler will extract the contract creation code (the binary of a smart contract), the ABI interfaces, and the constructor argument of those contracts. Furthermore, we also deploy the smart contracts on our Ethereum testnet. The deployed smart contracts serve two purposes: as subjects for fuzzing and as inputs for contract calls using contract address as argument.

The online fuzzing process begins with step 1 where the ContractFuzzer tool will analyze the ABI interface and the bytecode of the smart contract under test. Then it will extract the data types of each argument of ABI functions as well as the signatures of the functions used in each ABI function. In step 2, the tool will perform ABI signature analysis of all smart contracts crawled from Ethereum and then index the smart contracts by the function signature they support. This step is crucial for testing the interaction of smart contracts. Based on the analysis results on step 1 and step 2, the tool will generate valid fuzzing inputs conforming to ABI specification as well as mutated inputs across the border of validity in step 3. Note ABI not only can specify common data types as argument but also contracts address and functions selector as argument. The indexed smart contracts returned from step 2 are used to generate inputs for ABIs with contract address as arguments. Then in step 4, the tool will start the fuzzing process to bombard the generated inputs against the ABI interfaces with random function invocations. Finally, in step 5, the tool begins to detect security vulnerabilities by analyzing the execution logs generated during fuzzing. The fuzzing process continues until the available testing time is used up. The whole fuzzing campaign ends when the tool has finished fuzzing on all smart contracts under test.

## 4.2 Static Analysis of Smart Contracts

**Analysis of ABI Function Signatures of Contracts Pool**. The ContractFuzzer tool performs static analysis on ABIs the smart contracts pool to extract the signatures of the public functions supported by those contracts. More specifically, based on the exported ABI of each smart contract in JSON format, the tool extracts all function signatures declared in the ABI. We then calculate the function selector for each function signature, i.e., the first four bytes Keccak (SHA-3) hash of the signature of the function. Finally, we build a map with function selector as the key and a vector of addresses of all smart contracts having the same function selector as the value.

**Static Analysis of Smart Contract Under Test**. The Application Binary Interface (ABI) [15] of Ethereum smart contract is static and strongly typed. We parse the JSON format of the contract's ABI interface to extract its function descriptions and data types of each argument. The input domain of most data types can be determined precisely based on the documents [20]. However, the address data type essentially represents the address of contract account or external accounts. When the argument provided to the function is a contract address, the function may use the call function to interact with the contract. Therefore, when generating inputs for an ABI function with address type arguments, we must use the addresses of smart contracts that can support the functions called within the smart contract under test.

For a given ABI function of a smart contract under test, how can we effectively determine subset of smart contracts that it can interact with? The answer lies in the *call()* invocation within the bytecode of the function implementation. The first four bytes of the arguments for the *call()* function correspond to the first four bytes Keccak (SHA-3) hash of the signature of the function, which is also called the function selector.

Based on this observation, we perform static analysis of the smart contract bytecode to identify the function selectors used within the code of each public ABI functions as shown in Table 2.

The input of the algorithm is a smart contract in binary form and the output is a map, which maps each ABI function of the smart contract to a set of function selectors used in its code. At line 2, the algorithm first disassembles the smart contract bytecode into assembly code with the EVM tool *disam*. Then the algorithm extracts its set of public ABI functions as *funs* at line 4. From line 5 to line 16, there is a loop that iterates each function in *funs* to get the set of function selectors for it. Within the loop, it first extracts the body of the function and then locates the set of code segments of the function from line 6 to line 8. From line 10 to line 14, there is another loop that iterates each line of each code segment and checks whether the line starts with "PUSH4" opcode. If so, it will split the function selector. At line 15 and 16, the algorithm sets *M[f]* with the set of selectors. Finally, the algorithm returns the whole map M. The algorithm is performed on each smart contract in the pool.

**Table 2. The Function Selector Analysis Algorithm**

|    | |
|----|---|
|    | **function**:FindFunctionSelectorForABI |
|    | **input**: *bin*, smart contract in binary form |
|    | **output**: a map *M* recording the set of function selectors used in each ABI function of ABI |
| 1  | **def** FindFunctionSelectorForABI(bin): |
| 2  | dasm_bin = disamble(bin) //disassemble binary file |
| 3  | //extract each public ABI function signatures |
| 4  | funs = extractFunction(dasm_bin) |
| 5  | **foreach** f **in** funs: //iterate over each public function |
| 6  | body = extractBody(f) //extract body of function f |
| 7  | //get the code segments of body |
| 8  | codeSegs = getCodeSegments(body) |
| 9  | selector = set() |
| 10 | **foreach** seg in codeSeg: //iterate over code segments |
| 11 | **foreach** line in seg:  //iterate each line |
| 12 | **if** line.startswith("PUSH4") |
| 13 | //extract one function selector |
| 14 | selector.add(line.split()[1]) |
| 15 | **if** len(selector) >0: //ensure selector not empty |
| 16 | M[f]=selector |
| 17 | **return** M //return the map |

As discussed in previous sections, we have stored the addresses of all smart contracts with the same function selector in a map. So for each function selector returned from the function selector analysis algorithm, we search the map to find all smart contract addresses supporting that function selector as a ABI function. Then, we put those smart contracts into a private contracts pool for each ABI function of the smart contract under fuzzing. When generating fuzzing inputs for address data type of an ABI function, we will use the contract addresses within the private smart contract pool of the function, which makes the interactions of smart contracts possible.

## 4.3 Fuzzing Input Generation

**Input Generation Based on ABI Interface**. he input generation algorithm is responsible for generating valid inputs for each function. We use different strategies to generate inputs for fixed-size inputs and non-fixed-sized inputs. For fixed-size input types such as *INT<M>*, *UINT<M>*, *BYTES<M>*, and fixed array *<type>[M]*, we first build a set of values by randomly generating inputs within the valid input domain. Then we also build another set of seed inputs that are frequently used in smart contracts for that type based on static analysis. Then we combine these two sets to form the candidate value sets for that type. For non-fixed-size inputs such as *bytes*, *string*, and *<type>[]*, we generate the inputs in two steps. We first randomly generate a positive number as the length. Then we randomly select elements from their input domain.

The ContractFuzzer performs fuzzing for each function declared in the ABI of the smart contracts under test. Therefore, the input generation module aims at generating one set of candidate inputs for each function. To do so, the input generation module iterates each argument of the function to generate k candidates for each argument based on its input domain. Then the complete set of inputs is the combinations of k candidates of all arguments. Finally, the set of inputs are encoded into bytecode representation ready for invocation.

**Input Generation for Reentrancy Vulnerability**. Different from other vulnerabilities studied in this work, triggering the reentrancy vulnerability requires the interactive calls between two smart contracts. Therefore, we cannot expose such vulnerability by simply invoking a contract from an external account. Therefore, we need to generate a reentrant attack scenario to try triggering reentrancy vulnerability within the smart contract under test.

**Table 3. A Contract with Reentrancy Vulnerability**

| 1  | **contract** BountyHunt{ |
| 2  | ... |
| 3  | **function** claimBounty() preventTheft { |
| 4  | **uint** balance = bountyAmount[msg.sender]; |
| 5  | **if** (msg.sender.call.value(balance)()) { |
| 6  | totalBountyAmount -= balance; |
| 7  | bountyAmount[msg.sender] = 0; |
| 8  | } |
| 9  | } |
| 10 | } |
| 11 | contract AttackerAgent{ |
| 12 | ... |
| 13 | function AgentCall(address contract_addr,bytes msg_data){ |
| 14 | call_contract_addr = contract_addr; |
| 15 | call_msg_data = msg_data; |
| 16 | contract_addr.call(msg_data); |
| 17 | } |
| 18 | function() payable{ |
| 19 | call_contract_addr.call(call_msg_data); |
| 20 | } |
| 21 | } |

Therefore, we created an AttackerAgent contract to interact with each ABI function of the smart contract under test with a reentrant attack scenario. We use the testing of the *BountyHunt* smart contract as an example for illustration. The ContractFuzzer has also successfully detected reentrancy vulnerability within it in our experiment.

As shown in Table 3, the ContractFuzzer is fuzzing the *claimBounty* function of *BountyHunt* smart contract to determine whether it contains a reentrant bug. To do so, the fuzzer uses the *AttackerAgent* to try stealing ether from it with reentrant attack. At first, the *AgentCall* function of the *AttackerAgent* performs a call (line 14 to 16) to the *claimBounty* function of *BountyHunt* smart contract to initiate the attack (step 1). Within the *claimBounty* function, the *BountyHunt* send ether with a call at line 5 before setting the value of the corresponding bounryAccount to 0 at line 7. Since the call has no parameters provided, it will invoke the callback function (at line 18) of the *AttackerAgent* (step 2). Within the callback function, the *AttackerAgent* can invoke the *claimBounty()* function again to as the reentrant call (step 3). As a result, the *BountyHunt* will send ether to the AttackerAgent again until all of its ether is depleted. With the help of the AttackerAgent, we can ensure each reentrancy vulnerability flagged by ContractFuzzer is indeed exploitable.

### 4.4 Instrumenting EVM to Collect Test Oracles

Based on the definition of our proposed test oracles, we generally need to collect three types of information. The first type of information is about various attributes of a contract **call** or **delegatecall**. The second type of information is about the **opcodes** invoked during execution. The third type of information is the state of the contract during execution.

**Collecting Information on *Call/DelegateCall/Send*.** According to our definitions of smart contract vulnerabilities, all of them are related to the **Call/DelegateCall/Send** operation of a smart contract under fuzzing. The **Send** is treated as a special type of **Call** within EVM, And the **DelegateCall** is intrinsically the same as a **Call** except that the context and the storage of the caller rather than the callee is used during execution. Therefore, we can model all of them with the same **Call** data structure during instrumentation.

As shown in Table 4, for a **Call**, we collect the following information during its execution. These attributes of the call are crucial to support most of the test oracles.

**Table 4 Information Recorded for a Call**

| caller | address of the account that initiates the call |
|---|---|
| callee | address of the contract called |
| function | function called by caller |
| input | arguments for the function |
| value | amount of ether sent to callee |
| gas | gas stipend allowed for the call |
| internal_calls | calls made within the current smart contract |
| opCode_stack | opcode executed in the current call |

Furthermore, to support the test oracle **Reentrancy** and **ExceptionDisorder**, contract fuzzer requires information on cross contracts calls beyond the current contract under fuzzing. Therefore, we also record a chain of *Calls* invoked starting from the initial call, including the calls made in both the current contract and other contracts involved.

To record such information, we instrument of the *EVM.Call()* and *EVM.DelegateCall()* function within the EVM implementation of Ethereum to collect **Call** related information for each call. More specifically, we push the CALL opCode on to the opcode_stack, record information on internal calls, and also append the call onto the call chain.

**Collecting Information on Opcode.** Some test oracles such as **TimestampDependency** and **BlockNumDependency** must check the execution of some opcode such as TIMESTAMP and NUMBER. Moreover, many of the *opCodes* may change the state of the contract, which we will also need to record. To record opcode execution, we instrument the *interpreter.Run()* function within the Interpreter implementation of Ethereum. Upon the interpretation of each *opCode* monitored, we push it to the opCode stack of the current contract. We choose 34 out of the 129 EVM instructions for instrumentation in this work because they are either directly used by our test oracles or they may lead to the change of contract state, which are useful for security analysis.

### 4.5 Vulnerability Analysis and Report

The vulnerability analysis and report module is responsible for detecting the existence of any security vulnerability for report. When the call stack of the initial call becomes empty, the instrumentation module within the EVM will first checks those sub-oracles based on the collected instrumentation information. Then the instrumentation model will send those sub-oracles to a HTTP server listening at port 8888 of localhost. The server will collect the results of sub-oracles and then perform the final check of the composite conditions for each test oracle.

## 5 Experiment and Results Analysis

In this section, we present the details of our experiment as well as the results analysis.

### 5.1 Experiment Setup

We use a desktop PC as our testing environment. The PC is running Ubuntu 14.04 LTS and is equipped with Intel i5 8-core CPU and 16GB of memory. We configured two dockers within the PC to help setup the testing client and the Ethereum testnet. Within the docker running testing client, we used Ethereum Javascript API (web3.js libraries) within the node.js runtime to interact with the geth client within the testnet docker. The testnet docker installs the geth client version 1.7.0 and then we also created an Ethereum private blockchain within this docker with one peer node as the testnet. We set the initial mining difficulty of the genesis block with a very low value such that transaction confirmation can be fast during fuzzing. Finally, we deployed the 6991 smart contracts within this testnet for experimentation. Since we are focusing only on the smart contract vulnerabilities in this work, it is fine for our testnet to contain only one peer.

### 5.2 Smart Contracts as Subject Programs

There are about 9960 distinct smart contracts with verified source code on Etherscan [25] at the time of writing. We have crawled all of them and removed some smart contracts that cannot be deployed on our testnet due to their use of some invalid Ethereum address. Finally, the remaining 6991 smart contracts are used as our experimental subjects. We use the contract creation code (bytecode), the ABI interfaces, and the constructor argument of those contracts as the input for our ContractFuzzer. We choose these smart contracts with source code because they make it easier for us to manually verify the experiment results. But the ContractFuzzer only needs the bytecode of the smart contracts to work.

### 5.3 Experimental Procedure

The ContractFuzzer first performs static analysis on each contract to prepare the private contract pool for each ABI interface and to extract the ABI functions. With the static analysis result and the contract pool, the ContractFuzzer proceeds to generate inputs.

For each ABI function of each smart contract under fuzzing, ContractFuzzer call it with three types of account. The first one is an external account that is the creator the contract under test. The second one is a plain external account that has no relationship with the contract under test. The third one is a contract account called *AttackAgent* that can interact with the target contract with a reentrancy attack scenario. From each of the three types of account, we call the smart contract under test with two modes: one is with ether transfer and one is not.

For each ABI function, if contains arguments, ContractFuzzer will generate $k$ inputs to call it. Otherwise, we will simple perform one call on it. Combing with the 3 types of account and the 2 choices of call.value(), we will generate $6*k$ calls for a function with argument and 6 (3*2) calls for a function without argument. We initialize $k$ with a large value so that the ContractFuzzer can have a wide range of candidate calls to choose from. When fuzzing a specific smart contract, we merge all calls generated for each of its ABI function into a pool of calls to the smart contract. Then the ContractFuzzer starts the HTTP server to collect and analyze test oracles. For each smart contract, ContractFuzzer randomly selects calls from its pool of calls to perform fuzzing, which simulates different invocation sequences of the functions of the smart contract. Finally, the results are collected and analyzed by the server for report. We stopped the experiment after about 80 hours of fuzzing until the results gradually converge.

To compare our tool with the state of art smart contract verification tool, we also used the Oyente verification tool[9] to scan the 6991 subject smart contracts and compare the results with ContractFuzzer.

### 5.4 Experimental Results and Analysis

In this section, we present our experimentation results followed by detailed results analysis for each research question.

**Table 5 Summary of Vulnerability Detected**

| Vulnerability Type | Number of Vulnerabilities | Percentage | True Positive Rate |
|---|---|---|---|
| Gasless Send | 138 | 2.06% | 100% |
| Exception Disorder | 36 | 0.54% | 100% |
| Reentrancy | 14 | 0.21% | 100% |
| Timestamp Dependency | 152 | 2.27% | 96.05% |
| Block Number Dependency | 82 | 1.23% | 96.34% |
| Dangerous Delegatecall | 7 | 0.10% | 100% |
| Freezing Ether | 30 | 0.45% | 100% |
| Total | 459 | / | / |

**Summary of Vulnerabilities Detected.** In this section, we first summarize the results of our fuzzing campaign as shown in Table 5. The columns show the vulnerability type, the number of vulnerabilities found for the type, and its percentage over all smart contracts tested. The rows represent the results for each vulnerability type.

The number of smart contracts with *Gasless Send* vulnerability is 138, which is about 2.06% of all contacts tested. Since our test oracle for Gasless Send checks the occurrence of *ErrOutOfGas* error within the EVM, our results is precise and generates no false positives. For *Exception Disorder*, the ContractFuzzer detects 36 vulnerable smart contracts, which is about 0.54% of all smart contracts. We also manually checked each smart contract detected and confirmed that they were all true positives.

The number of *Reentrancy* vulnerability detected by ContractFuzzer is 14, which is about 0.21% of all smart contacts studied in the experiment. We checked these cases manually, and we confirm that there are no false positives reported. The infamous the DAO bug is also detected.

For *Timestamp Dependency* and *Block Number Dependency*, the ContractFuzzer detected 152 and 82 vulnerable smart contracts, respectively. We manually checked those smart contracts for confirmation. And we found 6 out of the 152 smart contracts flagged as Timestamp Dependency are false positives and 3 out of the 82 flagged as Block Number Dependency are false positives. Therefore, the true positive rates are 96.05% and 96.34%, respectively. The reason for the false positive cases is due to the imprecision of our test oracle definition for these two types of vulnerabilities. Indeed, in the test oracle definition for both cases, we checked the use of the opcode (TIMESTAMP and NUMBER) and the use of ether transfer calls within the ABI function tested. We have not checked whether there exists a dataflow def-use chain between the reading of the TIMESTAMP and NUMBER and the use of them in calculating the condition for transferring ether. However, recording such information may involve expensive data flow tracking through instrumentation of the smart contracts under test. Considering the high true positive rate, we consider the solution of ContractFuzzer is a cost-effective trade-off.

We detected 7 cases of *Dangerous Delegatecall* vulnerabilities, which we confirm were all true positives. These smart contracts all used the inputs of the initial caller to extract the functions to be called by *delegatecall*. The smart contract leading to the first round of Parity bugs are also detected. For the *Freezing Ether* vulnerability, we have detected 30 cases. We examined these smart contracts and confirm that all of them indeed have no way to send out ethers with their own code and they can only transfer ether indirectly through other contracts. In total, the ContractFuzzer has precisely detected 459 smart contracts as vulnerable with very high true positive rate.

**Table 6 Comparison of Contract Fuzzer and Oyente**

| Vulnerability Type | ContractFuzzer | | | Oyente | | |
|---|---|---|---|---|---|---|
| Statistics | No. | FP | FN | No. | FP | FN |
| Timestamp Dependency | 152 | 6 | 95 | 273 | 70 | 44 |
| Reentrancy | 14 | 0 | 1 | 43 | 28 | 0 |

**Comparison with the Oyente Verification Tool**. We also compared ContractFuzzer with the Oyente verification tool [9][29] on the detection of some type of vulnerabilities in common. The publicly released Oyente tool [29] can detect 4 types of vulnerabilities. One of them is the call stack size limit vulnerability that is no longer relevant since it is already addressed in the Ethereum EIP150 hardfork. The other 3 types of vulnerability detected by Oyente tool is the Transaction-ordering dependency, Timestamp Dependency, and Reentrancy. The testing of transaction-ordering dependency requires the manipulation of the transaction confirmation process, which is not yet supported by ContractFuzzer. So only the Timestamp Dependency and Reentrancy vulnerabilities can be detected by ContractFuzzer and Oyente in common. In general, if we add the transaction-ordering dependency vulnerability, ContractFuzzer can detect 7 out of 8 types of vulnerabilities while the Oyente can only detect 3 out of 8 types of vulnerabilities. So the ContractFuzzer can detect more types of vulnerabilities.

As shown in Table 6, for *Timestamp Dependency*, The ContractFuzzer detected 152 vulnerabilities and the Oyente detected 273 vulnerabilities. We examined those smart contracts manually and found both the ContractFuzzer and the Oyente have false positives and false negatives. The false negatives of the ContractFuzzer are due to two reasons. The first one is that some contracts hardcoded a specific time (i.e., a crowdfunding starting date) within their code and compares the block timestamp with it. As shown in Table 7, the code snippet for BDSM_Crowdsale smart contract specifies two fixed dates as the starting date and ending for ICO (line 3 and 4). Then the contract compares the block timestamp with the ICO starting dates to determine whether to perform a refund (line 6 and 7). Obviously, the smart contract is timestamp dependent since its uses the current block timestamp to decide whether to perform ether transfer. However, since the specific date has passed already, the condition will always fail. As a result, no ether transfer will happen during execution and ContractFuzzer will not consider it timestamp dependent. The second reason for the false negatives of ContractFuzzer is some conditions are hard to trigger with limited testing time. For example, one contract requires a specific call pattern of the functions before triggering a timestamp dependent ether transfer. We can perform more extensive fuzzing with different function call sequences to improve this situation.

**Table 7. A False Negative Case of ContractFuzzer**

| 1 | **contract** BDSM_Crowdsale { |
|---|---|
| 2 | ... |
| 3 | uint public startICO_20_December = 1513728060; //2017.12.20 |
| 4 | uint public stopICO_20_March = 1521504060; //2018.3.20 |
| 5 | function () payable { |
| 6 |     if (now < startICO_20_December) { |
| 7 |         msg.sender.transfer(msg.value); |
| 8 |     } |
| 9 | ... |

The false negatives of the Oyente tool are mainly due to the difficulty of symbolically analyzing cryptographic operations. Many of the vulnerable smart contract makes use of the block timestamp or the block number as the source for random seed generation with cryptographic function. For example, the code snippet from the *Bomb* contract at Table 8 invokes cryptographic functions such as blockhash and keccak256 with block number and timestamp as input (line 3). And then it checks whether the results is equal to 1 (line 4) before transferring ether. This code is almost unfeasible to be symbolically analyzed, as it is basically asking for an input that inverts 1, which is exactly what cryptographic function is designed to make difficult. As a result, the Oyente may miss the vulnerabilities written with this pattern. There are also some false positive cases of Oyente tool. After manual inspection, we found the timestamp is never used to calculate the path condition to transfer ether.

For *Reentrancy Vulnerability*, The ContractFuzzer detected 14 vulnerabilities and the Oyente detected 43 vulnerabilities. With manual examination, we confirmed all 14 vulnerabilities detected by ContractFuzzer are all true positives. However, there is one vulnerable smart contracts missed by for ContractFuzzers (i.e., false negatives). We checked these smart contracts and found that the vulnerable function of some smart contracts must perform complex condition checks before transferring ether. However, these conditions are hard to trigger by ContracFuzzer.

Table 8. A False Negative Case of Oyente

| 1 | **function** buy(uint8 _bomb) **public payable** { |
|---|---|
| 2 | ... |
| 3 | **int** _random = **uint**(**keccak256**(block.**blockhash**(block.**number**-1),msg.gas,tx.gasprice,block.**timestamp**))%bomb.chance + 1; |
| 4 | **if**(_random == 1){ |
| 5 | bomb.owner.**transfer**(...) |
| 6 | ceoAddress.**transfer**(...) |
| 7 | }} |

For the Oyente tool, 28 out of the 43 smart contracts flagged as reentrant are false positives with our test oracle definition. We classified the 28 false positive cases into 3 types. The first type is the smart contract uses *send()* and *transfer()* operation with a limited gas stipend which makes the callback function of the callee have no enough gas to perform the reentrant call again. The vulnerable functions of the second type of smart contracts strictly check whether the caller of them is the owner of the smart contract specified during contract creation. Since there is no way for an external account to invoke the function containing ether transfer, reentrant attack is also not possible. Finally, the third type of smart contracts can only transfer ether to a fixed address hard-coded within the smart contract. A malicious contract has no way to get the ether or triggering reentrant call. Therefore, these 28 cases are false positive cases with our definition: reentrancy can never be triggered from an external contract. In contrast, the ContractFuzzer did not report those false positive cases because it ensured that both the reentrant call and the transfer of ether to external account did happen.

To conclude, when compared with Oyente, the ContractFuzzer has much lower false positive rate on both vulnerabilites. However, the false negative rates of ContractFuzzer is high for *Timestamp Dependency*, which we may use more comprehensive input generation schemes to improve.

## 5.5 Case Studies on Attacks on Vulnerable Smart Contracts

In this section, we will present cases studies on some exploitable smart contracts flagged as vulnerable by *ContractFuzzer*.

**Wrongfully Holding Ether of Investors**. The CrowdSalePreICO in Table 9 is a malicious smart contract that makes use the exception disorder vulnerability to wrongfully hold the ether of the investors. The function at line 3 is the callback function of the smart contract. The ContractFuzzer will call the CrowdSalePreICO with an empty input and some value to serve as an investor. Then the callback function will check the received ether and add it to the total deposit (line 5 to 8). However, if the total deposit exceeds the crowd sale limit, the excessive ether will be returned to the investor (line 9 to 11). However, the send at line 11 may fail. But the CrowdSalePreICO smart contract will not check and handle the error. As a result, the fundraiser will wrongfully keep the excessive ether of the investor.

Table 9. ICO Contract Wrongfully Holding Ether of Investors

| 1 | **contract** CrowdSalePreICO { |
|---|---|
| 2 | ... |
| 3 | **function**() payable stopInEmergency onlyStarted notFinalized{ |
| 4 | ... |
| 5 | uint contribution = msg.value; |
| 6 | **if** (safeAdd(totalDepositedEthers, msg.value) >hardCapAmount) { |
| 7 | contribution = safeSub(hardCapAmount, totalDepositedEthers); |
| 8 | } |
| 9 | uint excess = safeSub(msg.value, contribution); |
| 10 | **if** (excess > 0){ |
| 11 | msg.sender.send(excess); |
| 12 | }}} |

**Manipulating Timestamp to Win Slot Machine**. As shown in Table 10, the contract SlotMachine() is a smart contract that makes use of the block timestamp as random number for determining the winner of the slot machine game.

Table 10. An Exploitable Slot Machine Smart Contract

| 1 | **contract** SlotMachine { |
|---|---|
| 2 | ... |
| 3 | **function**(){ |
| 4 | uint nr = **now**; //now is the block timestamp |
| 5 | uint y = nr & 3; |
| 6 | ... |
| 7 | **if**(y==1) { wins[1]++; win = (msg.value * 2) + (msg.value / 2);} |
| 8 | earnings += int(msg.value); |
| 9 | **if**(win > 0) { |
| 10 | bool res = msg.sender.send(win); |
| 11 | earnings -= int(win); |
| 12 | }}} |

At line 4, the block timestamp is read into nr. Then nr is used to calculate win (line 5 to 7). Then the win is used to determine whether send the reward to the caller of the function (line 9 to 11). If a miner of the Ethereum blockchain participates the slot machine game, he can manipulate the value of block timestamp (i.e. now) in favor of he/her interest. This kind of attack exists in all smart contracts that makes use of the block timestamp and block number to determine the condition for ether transfer.

**Relying on Hard-Coded Library to Transfer Ether.** As shown in Table 11, which is the wallet contract affected by the second round of parity bug. Within the Wallet smart contract, it can invoke the code of walletLibrary to manipulate its account or perform other operations (line 7 and line 10). However, the code of Wallet contains no call/transfer/suicide to transfer ether. What is worse, the walletLibrary is defined as a hard-coded address. When walletLibrary is changed to a smart contract account and killed, the Wallet contract has no way to send out ether and its ether is frozen. During the attack of the parity vulnerability on Nov. 2017, $280M ether was frozen within such Wallet smart contracts account.

**Table 11. The Wallet Smart Contracts Frozen by Parity Bug**

| 1  | contract Wallet is WalletEvents { |
| --- | --- |
| 2  | ... |
| 3  | function() payable { |
| 4  |   if (msg.value > 0) |
| 5  |     Deposit(msg.sender, msg.value); |
| 6  |   else if (msg.data.length > 0) |
| 7  |     _walletLibrary.delegatecall(msg.data); |
| 8  | } |
| 9  | function hasConfirmed(bytes32 _operation, address _owner)...{ |
| 10 |   return _walletLibrary.delegatecall(msg.data); |
| 11 | } |
| 12 | address constant _walletLibrary = 0xcafecafecafecafecafecafecafecafecafecafe; |
| 13 | } |

# 6 Related Work

In this section, we briefly review related works on smart contract vulnerabilities and attacks as well as security testing and verification techniques to detect such security bugs.

## 6.1 Smart Contracts Bugs and Vulnerabilities

Delmolino et al. [4] showed that there could be a lot of logical problems in even a tiny smart contract. And they listed some common logical problems such as contract never refunding ether to its sender and non-encrypted data leaking privacy. Miller [10] audited the source code of smart contracts, and reports call-stack overflow bugs on Ethereum. After undergoing continuous attacks in 2016, the problem is resolved via a hard fork on Ethereum. Atzei [1] systematically surveyed security attacks on Ethereum smart contracts. They provided taxonomy of smart contract vulnerabilities based on their characteristics.

## 6.2 Smart Contracts Security

Fillâtre [6] presented Why3, a tool used for program verification, which is now available within Solidity Web IDE [9] as the formal verification backend. When programming smart contract within the IDE, the tool can help check the arrangement of integer array, the overflow of mathematic operation and the division by zero bugs.

Bhargavan [3] managed to study the security of smart contracts and the reliability of Solidity compiler. They designed a tools suite, which could transform both Solidity source code and EVM bytecode to F* [3] program, respectively. Luu et al [9] designed Oyente, a symbolic verification tool for smart contract. Oyente builds the control-flow graph of smart contracts and then performs symbolic execution on the control flow graph while checking whether there exist any vulnerable patterns.

Nikolic et al [12] designed MAIAN, a symbolic execution tool for reasoning about tracing properties to detect vulnerable smart contracts. It specified three typical smart contracts vulnerabilities based on trace properties. The MAIAN can efficiently detect the greedy, the prodigal and the suicidal contracts through symbolic execution. Different from their work, the ContractFuzzer tool performed fuzzing and runtime monitoring to detect vulnerabilities that happened during execution, which can generate fewer false positives. Hirai [8] used Isabelle/HOL tool to verify the smart contract called Deed, which is a part of Ethereum Name Service implementation. Specifically, the work verifies the oracle that only the owner of Deed could decrease its balance. Furthermore, they also found the EVM implementation is poorly tested during the verification process. Echidna [21] is a smart contract fuzzer with oracles defined within unit test by testers. It can also generate inputs to fuzz smart contracts. However, it provides no direct API supporting security testing of smart contracts.

## 6.3 Fuzzing Techniques for Vulnerability Detection

Many of the black-box fuzzers are grammar-based such as SPIKE [2] and Peach [5]. Hanford [7] and Purdom [13] started to study test case generation based on grammar in 1970s. They showed grammar-based fuzzers are effective to detect vulnerabilities within the application under testing.

# 7 Conclusion

With the popularity of blockchain and smart contract technique, millions of smart contracts have been deployed on blockchain platforms to enable the building of decentralized applications. However, the security vulnerabilities of the smart contracts pose big threat to their future. In this work, we proposed ContractFuzzer, a precise and comprehensive fuzzing framework to detect 7 types of Ethereum smart contract vulnerabilities. Our experiment with 6991 real world smart contracts shows that the input generation and test oracle analysis strategies of ContractFuzzer are effective to trigger and detect security vulnerabilities with very high precision. Our tool reported 459 vulnerabilities in total out of the 6991 smart contracts tested, including the infamous the DAO bug and the Parity Wallet bug. When compared with the state of art security verification tool Oyente, the ContractFuzzer not only can detect more types of vulnerabilities but it also has much lower false positives.

For future work, to reduce false negatives, we may study the vulnerability exploit patterns for those types of smart contract bugs, which may guide us to generate more effective vulnerability triggering inputs. We will also extend our tool to detect more types of smart contract vulnerabilities related to the EVM or the underlying blockchain platform. Finally, we will also generalize our work to the security testing of other smart contract platforms.